\documentclass[useAMS,usenatbib,usegraphicx]{mn2e}
\usepackage{psfig}   
\usepackage{graphicx}
\usepackage{subfigure}

\newcommand{\Msun}{\ensuremath{\textrm{ M}_{\odot}}}

\newcommand{\nth}{\ensuremath{^{\rm th}}~}

\title[On the formation of trapezium-like systems]{On the formation of
  trapezium-like systems}

\author[R.~J.~Allison et al.]{ Richard~J.~Allison$^1$, Simon
  P.~Goodwin$^1$\thanks{E-mail: s.goodwin@sheffield.ac.uk}
  \vspace*{0.1cm}\\ $^1$ Department of Physics and
  Astronomy, University of Sheffield, Sheffield, S3 7RH, UK\\}

\begin{document}

\date{}
                             
\pagerange{\pageref{firstpage}--\pageref{lastpage}} \pubyear{2010}

\maketitle

\label{firstpage}

\begin{abstract}

We investigate the formation and evolution of high-order massive star
multiples similar to the Trapezium in the Orion Nebula Cluster.  We
perform ensembles of $N$-body simulations of the evolution of $N=1000$
Orion-like clusters with initial conditions ranging from cool and
clumpy to relatively smooth and relaxed.  We find that trapezium-like
systems are frequently formed in the first 2~Myr in initially cool and
clumpy clusters and can survive for significant amounts of time in
such clusters.  We also find that these systems are highly dynamical
entities, constantly interacting with the surrounding cluster,
changing their appearance and membership regularly.  The eventual
decay of trapezium-like systems can even destroy the host cluster.  We
argue that the current state of any trapezium-like system is transient
and care should be taken when analysing and drawing conclusions from a
single snapshot in the life of a highly dynamic object.

\end{abstract}


\section{Introduction}
\label{sec:intro}

Stars in our local environment (i.e. within 500 pc of our Sun)
form in a continuous hierarchical distribution \citep{bressert10} and
dynamically evolve to form over-dense groups of stars that we call
`clusters' \citep[e.g.,][]{gutermuth05,portegies_zwart10}. The star
formation process is thought to be driven by turbulence which imprints
substructure in the clouds, thus star formation occurs in clumpy and
filamentary regions of molecular clouds
\citep[e.g.,][]{maclow04,ballesteros-paredes07}. Observations of
molecular clouds have show that star formation in clumps and filaments
is the norm in molecular clouds
\citep[e.g.,][]{testi00,gutermuth05,sanchez07,andre07,goldsmith08,andre10,difrancesco10},
a result which is also seen in hydrodynamic simulations
\citep[e.g.,][]{klessen00,bonnell01,bonnell03,bate03,bonnell08,offner09}. 
Observations suggest that these newly born stars have sub-virial
velocities
\citep[e.g.,][]{walsh04,difrancesco04,peretto06,walsh07,andre07,kirk07,gutermuth08b},
a result which is also found in simulations
\citep[e.g.,][]{klessen00,bonnell01,bonnell03,bate03,bonnell08,offner09}.
Dynamical simulations have also shown that initially
 subvirial conditions are able to reproduce observations of the
 Orion star forming region
 \citep{adams06,proszkow09,allison09,allison10}, and 
 erase initial substructure on short timescales
 \citep{goodwin04,allison09,allison10}.

Both observations and theory indicate that the dynamical evolution of
star clusters begins with clumpy and subvirial initial conditions. An
early attempt to model young clusters (i.e., those with substructure)
was made by \citet{aarseth72}, who simulated an initially
structured cluster using a linear distribution of clumps with
velocities set to be at rest. In this early model it was found that
the dynamical evolution of the system was very rapid, and that the
structure was destroyed within its first crossing time. It was also
found that the formation of binary systems was increased and that
stars were more likely to be ejected, and at higher velocities, with
respect to a non-structured system. It was also found that initial
structure could lead to the formation of high-order multiple systems
\citep[such as the Trapezium system in the ONC,][]{aarseth77}.

Following \citet{allison10} we investigate the ensemble of initially
substructured and subvirial cluster simulations to
study the formation and evolution of high-order, high-mass multiple
systems; which we define as `trapezium-like systems' after the
canonical Trapezium system in the Orion Nebula Cluster (ONC)
\citep{ambartsumian54}. In these ONC-like clusters, trapezium-like
systems contain the highest mass stars in the cluster, and as such
have the potential to dominate the dynamical evolution of the host
cluster. In this paper we investigate how and when these
systems form dynamically and the affect these systems have on their
host clusters. In Section~\ref{sec:init} we describe our initial
conditions. In Section~\ref{sec:define_trap} we define what we
consider to be a trapezium-like system, and how our detection
algorithm finds multiple star systems. In Section~\ref{sec:dyn} we briefly
discuss the results presented in \citet{allison10} and in
Section~\ref{sec:tresults} we present our results for this work. In
Sections~\ref{sec:discussion} and~\ref{sec:tconc} we discussion the
implications of this work and draw our conclusions.

\section{Initial Conditions}
\label{sec:init}

We perform 160 $N$-body simulations of cool, clumpy star clusters.  We
vary the level of substructure and initial virial ratio. We conduct
ensembles of simulations with statistically identical initial
conditions, varying only the initial random number seed used to
initialise the simulations. These simulations are the same as are
described in
\citep{allison09,allison10}, but are reiterated here for clarity.

To create initial substructure in our simulations we use a fractal
stellar distribution. Using a fractal distribution provides a
parameterisation of substructure using only a single number: the
fractal dimension.  (Note that we are not claiming that clusters are
actually initially fractal, although they may be, just that this
provides a simple descriptor of substructure that is easy to
produce.)

The fractal stellar distributions were generated following the
method of \citet{goodwin04}. The method begins by defining a cube
of side $N_{{\rm div}}$, inside of which the fractal will be built. A
first-generation parent is placed at the centre of the cube, from
which are spawned $N_{{\rm div}}^{3}$ sub-cubes, each containing a
first-generation child in its centre. The fractal is then built
by determining which of the children themselves become parents,
and spawn their own offspring. This is determined by the fractal
dimension, $D$, where the probability that a child becomes a
parent is $N_{\rm div}^{(D-3)}$. For a lower fractal dimension
less children will mature and so the final distribution will
contain more structure. Any children which do not become parents
in a given step are removed, along with their parent. A
small amount of noise is then added to the positions of the
remaining children, preventing the final cluster from having
a gridded appearance, and the children become parents of the next
generation. Each new parent then spawns $N_{\rm div}^{3}$
second-generation children in $N_{\rm div}^{3}$ sub-sub-cubes,
with each second-generation child having a $N_{\rm div}^{(D-3)}$
probability of becoming a second-generation parent. This process
is then repeated until there are substantially more children than
required. The children are pruned to produce a sphere from the
cube and are then randomly removed (so maintaining the fractal
dimension) until the required number of children are left. These
children then become the stars in the cluster.

To determine the velocity structure of the cloud, children inherit
their parent's velocity plus a random component that decreases with
depth in the fractal.  The children of the first generation are given
random velocities from a Gaussian of mean zero. Each new generation
then inherits their parent's velocity plus an extra random component
that becomes smaller with each generation.  This results in a velocity
structure in which neighbour stars have similar velocities, but distant
stars can have very different velocities.  Finally, the velocity of
every star is scaled equally to obtain the desired total virial ratio for the
cluster.

\begin{center}
\begin{table}
  \begin{tabular}{|c|c|c|c|c|}
\hline
        &  \multicolumn{4}{|c|}{$D$} \\
$Q$     &  1.6        &  2.0        &  2.6        &  3.0  \\
\hline
0.3     &  a1.01--50  &  a2.01--10  &  a3.01--10  &  a4.01--10   \\
0.4     &  b1.01--10  &  b2.01--10  &  b3.01--10  &  b4.01--10   \\
0.5     &  c1.01--10  &  c2.01--10  &  c3.01--10  &  c4.01--10   \\
\hline
  \end{tabular}
\caption{Notation for run identification where $D$ is the initial
  fractal dimension, and $Q$ is the initial virial ratio of each
  simulation. Within each ensemble only the random number seed used to
  generate the initial conditions is changed.
\label{tab:runs}}
\end{table}
\end{center}

Each simulation contains 1000 stars, has an initial maximum radius of
1 pc, includes no primordial binaries or gas, and has a three-part power law
is used to produce an initial mass function \citep[IMF,][]{kroupa02},

\begin{equation}
  N(M) \propto \left\{
  \begin{array}{r}
    M^{-0.3} \quad m_0 \leq M/\Msun < m_1, \\
    M^{-1.3} \quad m_1 \leq M/\Msun < m_2, \\
    M^{-2.3} \quad m_2 \leq M/\Msun < m_3, \\ 
  \end{array}
  \right.
\end{equation}

\noindent with $m_0=0.08\Msun$, $m_1=0.1\Msun$, $m_2=0.5 \Msun$ and
$m_3=50\Msun$. No stellar evolution is included because of the short
duration of the simulations ($\sim 4$ Myr).  We use the {\sc starlab}
$N$-body integrator {\sc kira} to run our simulations
\citep{portegies_zwart01}.

In this study we explore a range of fractal dimensions and virial
ratios. The fractal dimensions investigated are
$D=1.6,~2.0,~2.6~\textrm{and}~3.0$ (since these values correspond to
the number of maturing children, $2^{D}$ , being an integer), where
$D=1.6$ produces a large amount of structure and $D=3.0$ produces a
roughly uniform sphere. We investigate virial ratios of $Q =
0.3,~0.4~\textrm{and}~0.5$, we define the virial ratio as
$Q=T/|\Omega|$ (where $T$ and $|\Omega|$ are the total kinetic and
total potential energy of the stars, respectively), hence virial
equilibrium is $Q=0.5$. The `sets' of initial conditions are tabulated
in Table~\ref{tab:runs}.

It is important to note that fractal initial conditions are inherently
stochastic: statistically identical fractals (i.e., the same fractal
dimension) can appear very different to the eye, and can evolve in
very different ways \citep[][]{allison10}.  Therefore, it is vital to
perform ensembles of simulations with different random number seeds.
We have therefore simulated 50 $D=1.6$, $Q=0.3$ (a1) and $D=1.6$,
$Q=0.4$ (b1) clusters (as they have the most interesting evolution,
and to investigate anomalous results caused by low number statistics),
and restricted our analysis of all other combinations of $D$ and $Q$
to 10 clusters each.

The trapezium-like systems described in this paper have been formed from
the collapse and subsequent evolution of clusters, they are {\em not} 
primordial systems.  Indeed, the high-mass stars that end-up
in a system may have started a significant distance apart from
each-other.  

\section{What is a `trapezium-like system'?}
\label{sec:define_trap}

 The classical idea of a trapezium system is the multiple star
  system in the ONC - The Orion Trapezium system. The Trapezium system
  has a very complicated layout, comprising the four main OB-stars
  ($\theta^1$~Ori A, B, C and D) and their binary components
  \citep{preibisch99}. $\theta^1$~Ori B is in fact a `mini-trapezium'
  on its own and has at least four lower-mass companions
  \citep{weigelt99}, and $\theta^1$~Ori A and C are known to have
  binary companions. The brightest star in the system is
  $\theta^1$~Ori C whose companion star has a mass approximately half
  its own mass (making this a O-star -- O-star or OB-star binary
  system), this binary has a relatively short period of $\sim
  11$~years \citep{kraus07}. Thus the `classical' idea of the four
  OB-star Trapezium system is a misnomer, as the Trapezium system is
  at the least an $N = 8$ system, if we ignore very close
  binaries \citep[see, e.g., the schematic of the Trapezium;][their
    Figure~9]{moeckel09b}. If $\theta^1$~Ori E is included as a member
  of the Trapezium system it would become an $N \sim 9$ system, as this
  star is also a close binary system \citep{zinnecker07}.  Therefore,
  we use the term `trapezium-like' (or simply `trapezium') to identify
  any multiple-star system that contains several O-stars with lower-mass
  companions.

We define a trapezium system as a system of at least three high-mass
stars (with possibly other massive stars and other low-mass stars
present) that are `observed' to be in a multiple system.  These
systems may not be gravitationally bound, although that some systems
survive for extended periods of time does indicate that most systems
we detect are bound in some sense.  This differs from the usual
definition, i.e. a trapezium system is a multiple star system whose
pairwise separations are of the same order
\citep[e.g.,][]{ambartsumian54,pflamm-altenburg06}. While we have
chosen to use a definition different from this, the method we use
allows much useful information about any detected system to be
gathered; such as the minimum spanning tree of the system, and a
complete list of binary stars in the cluster. This extra information
allows an in-depth study of the systems we find.

Our multiple system finding algorithm starts by finding the
  nearest and second nearest neighbours of each star.  It also finds
  the local density from the volume enclosing the 40\nth nearest
  star\footnote{This is a larger volume than is usually considered,
    which would normally only contain a handful of stars
    \citep[see][p. 265]{aarseth03}.  This is because we want a density
    estimator that includes all of the members of a high-order
    multiple system and surrounding stars as well.}.  Higher-order
  multiples are found by looking for closed loops of first and second
  nearest neighbours in several steps.

First, mutual nearest neighbours are linked together as long as their
separations are less than one-third of the typical separation for the 
local density.  (We note that the algorithm is not very sensitive to
changes such as using the 30\nth nearest neighbour and one-half of the
typical local separation.)  Pairs of mutual nearest neighbours are
then used as a starting point to search for higher-order multiple
systems, and both members of the pair are registered as being
potential members of a system.

In the second step, the nearest and second nearest neighbours of each
potential member are 
examined and registered as being potential members of a system if
their separations are less than one-third of the typical local
separation (some might already be potential members, others may be new).  

The third step is to see if the potential members form a closed loop.
That is: are all of the nearest and second nearest neighbours (within
one-third of the typical local separation) of potential members, also
potential members of the system?  If so we have a closed loop.

Finally, if a closed loop is not found we return to the second step
and look for new potential members.  Numerical experiments show that
if a closed loop is not found in a few iterations (normally four or
five) then a higher-order multiple does not exist.  Therefore we
restrict ourselves to eight iterations.

For example, the simplest higher-order system is a triple system.  If
stars $i$ and $j$ are mutual nearest neighbours then they are the
starting point for the search.  If they both have the same second
nearest neighbour, $k$, then this might be a triple system.  It is a
triple system if the nearest and second nearest neighbours of $k$ are
stars $i$ and $j$ (we have a closed loop).  However, if one of the
neighbours of $k$ is not $i$ or $j$ then this new star becomes a
potential member and we return to the second step.

There is one slight subtlety as outlying members of higher-order
systems can be missed by this method.  For example, stars $i$ and $j$
are both members of a higher-order system with a closed loop.  Star
$k$ has $i$ and $j$ as its nearest and second nearest neighbours --
therefore it should be part of the system.  However, it can happen
that no members of the loop have $k$ as a neighbour and therefore it
is missed.  To avoid this problem a final step is to check if any
stars have both neighbours as members of the system but have not been
included themselves.

As a very final step, trapeziums are found by searching for
higher-order multiples that contain at least three high-mass 
stars $>8$~M$_\odot$ \citep[e.g.,][]{abt00}.  It should be noted
  that in our definition of a trapezium a hard massive star-massive
  star binary orbited by a third massive star would be identified as a
  `trapezium'.  However, as we show in the results (in particular
  Figure~2), most of our `trapeziums' are higher-order systems that
  we feel most people would agree are `trapezium-like').

We note that this method can be applied in two- or three-dimensions
and the boundness of the system in three-dimensions can be found to locate
`real' higher-order multiples.  

\section{The dynamical evolution of cool, clumpy clusters}
\label{sec:dyn}

In this section we briefly discuss the work presented in
\citet{allison10}, which is a precursor to the work presented here. 

\citet{allison10} show that clusters with clumpy and cool initial
conditions dynamically evolve in a significant way on 
very short timescales ($\sim 1$~Myr), and
that this rapid evolution allows the clusters to {\em dynamically} 
mass segregate at
young ages. The initial conditions of cool and clumpy clusters places
the clusters far from an equilibrium state, causing the cluster to
enter a phase of violent relaxation \citep{lynden-bell67}. During the
violent relaxation phase the cluster collapses, erasing its initial
substructure, and approaches virial equilibrium. The collapse also
causes the formation of a dense core which allows rapid dynamical
mass segregation to occur. The cooler and more substructured the
cluster is initially, the more it will collapse as it attempts to
relax and reach virial equilibrium. This is shown by
Eq~\ref{eq:collapse} \citep[see also,][]{allison10};

\begin{equation}
\label{eq:collapse}
  \frac{R_0}{R_f} = \frac{\eta_0}{\eta_f} 2(1 - Q_0),
\end{equation}

\noindent where $R_0$ and $R_f$ are the initial and final radii of the
system, respectively; $\eta_0$ and $\eta_f$ are the initial and final
structure parameters, respectively; and $Q_0$ is the initial virial
ratio. The value of $\eta$, the structure parameter, depends on
  the structure of the cluster, it is therefore a measure of the
  distribution of potential energy in the cluster
  \citep{portegies_zwart10}. The more substructure the cluster has
initially the greater $\eta_0$, and for a substructured cluster
$\eta_0>\eta_f$. This collapse leaves the cluster with a 
very short relaxation time in the core allowing mass
segregation to occur on timescales much shorter than usually expected
\citep[e.g.,][]{bonnell98}. The initial conditions of the cluster
influences the ability of the cluster to mass segregate by determining
the depth of the collapse of the cluster -- the cooler and clumpier the cluster
initially the denser the core it will form.

\section{Results}
\label{sec:tresults}

In this section we analyse and discuss the properties of the trapezium
systems found in our simulations.

\subsection{The nature of trapezium systems}

\begin{figure*}
  \begin{center}
    \setlength{\subfigcapskip}{10pt}
\subfigure[]{\label{sfig:faa11a011.dat_trap}
  \includegraphics[scale=0.35,angle=270]{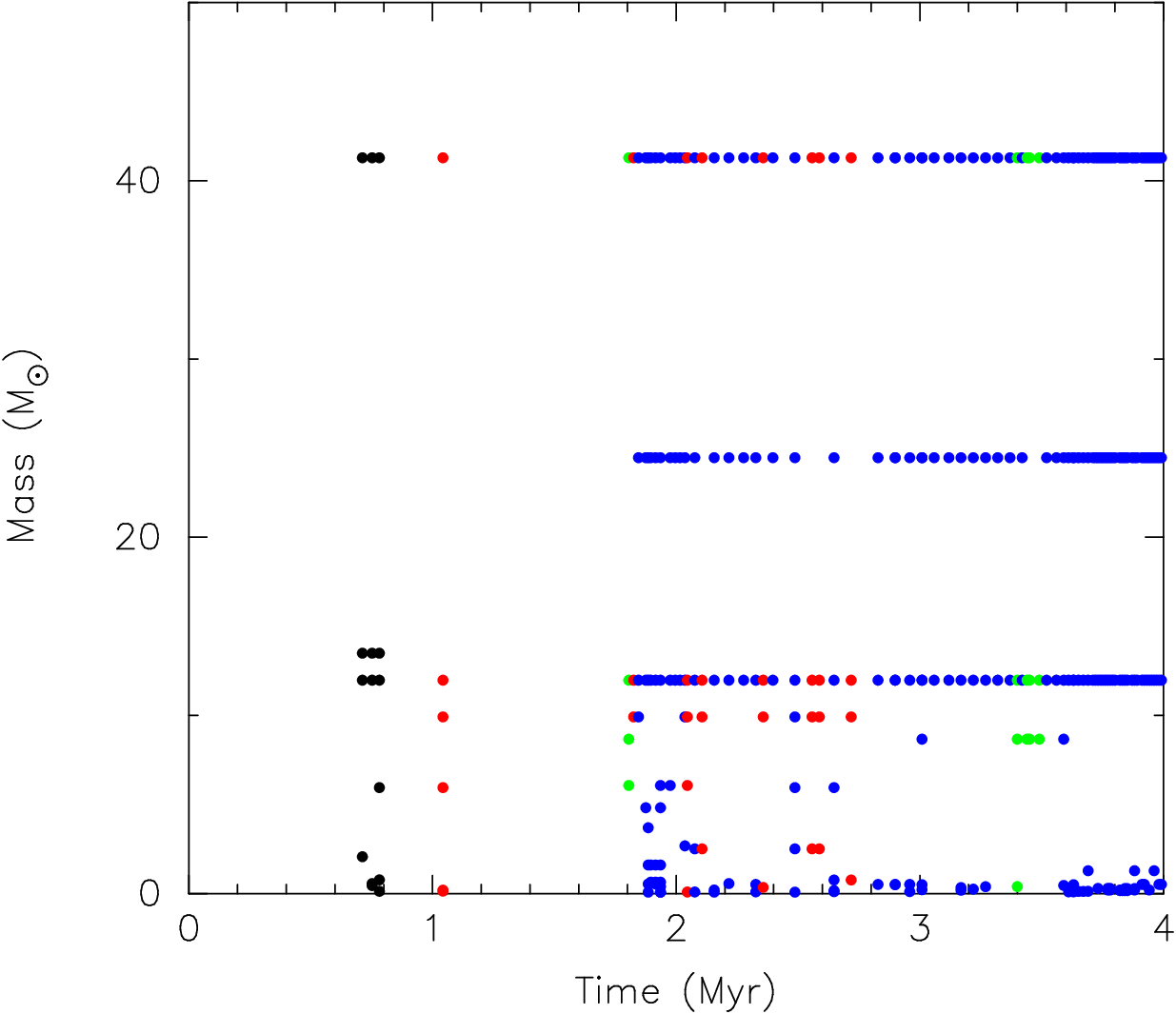}}
\subfigure[]{\label{sfig:faa11a011.dat_trap.071}
  \includegraphics[scale=0.35,angle=270]{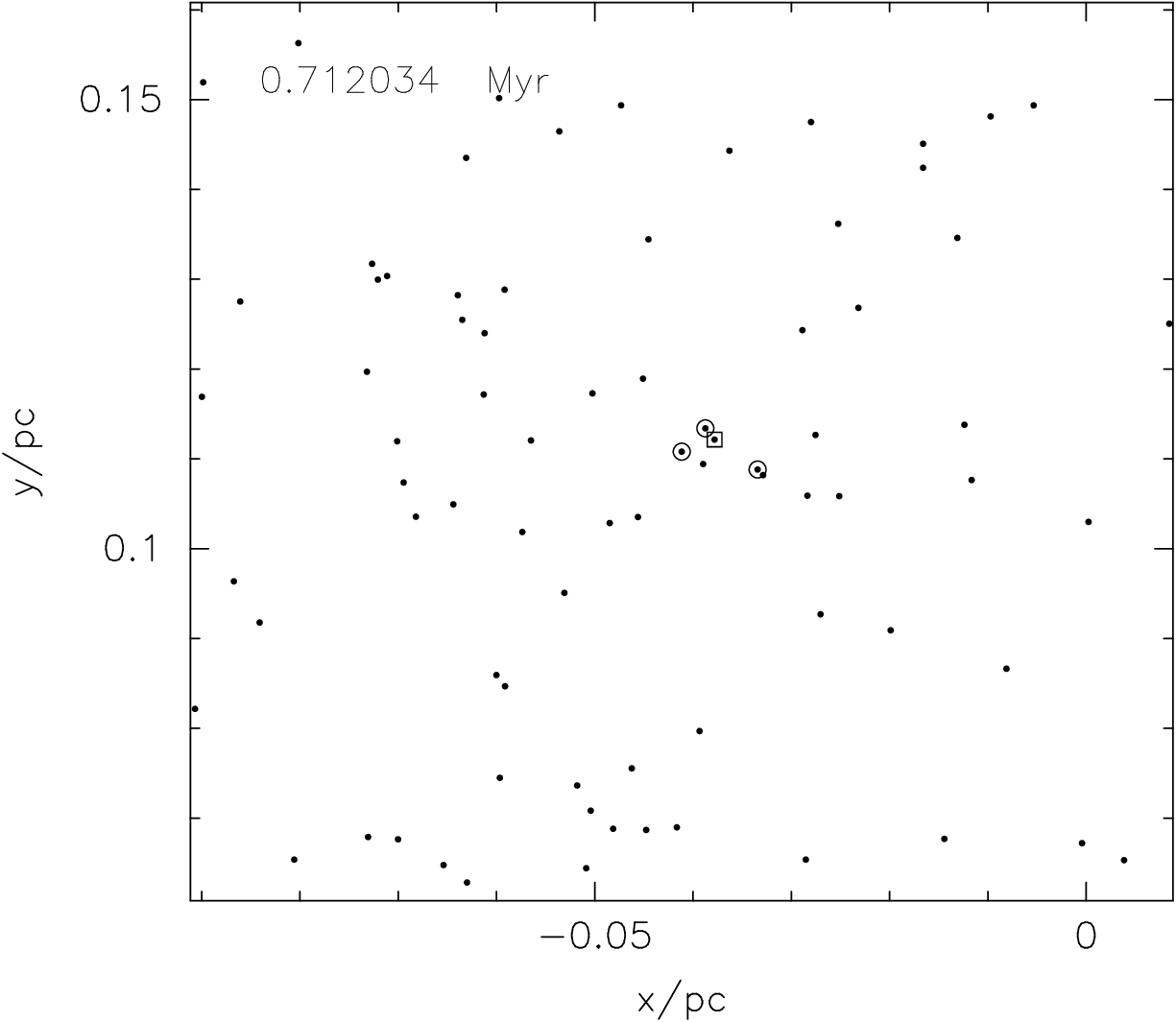}}
\vspace{10pt}
\subfigure[]{\label{sfig:faa11a011.dat_trap.090}
  \includegraphics[scale=0.35,angle=270]{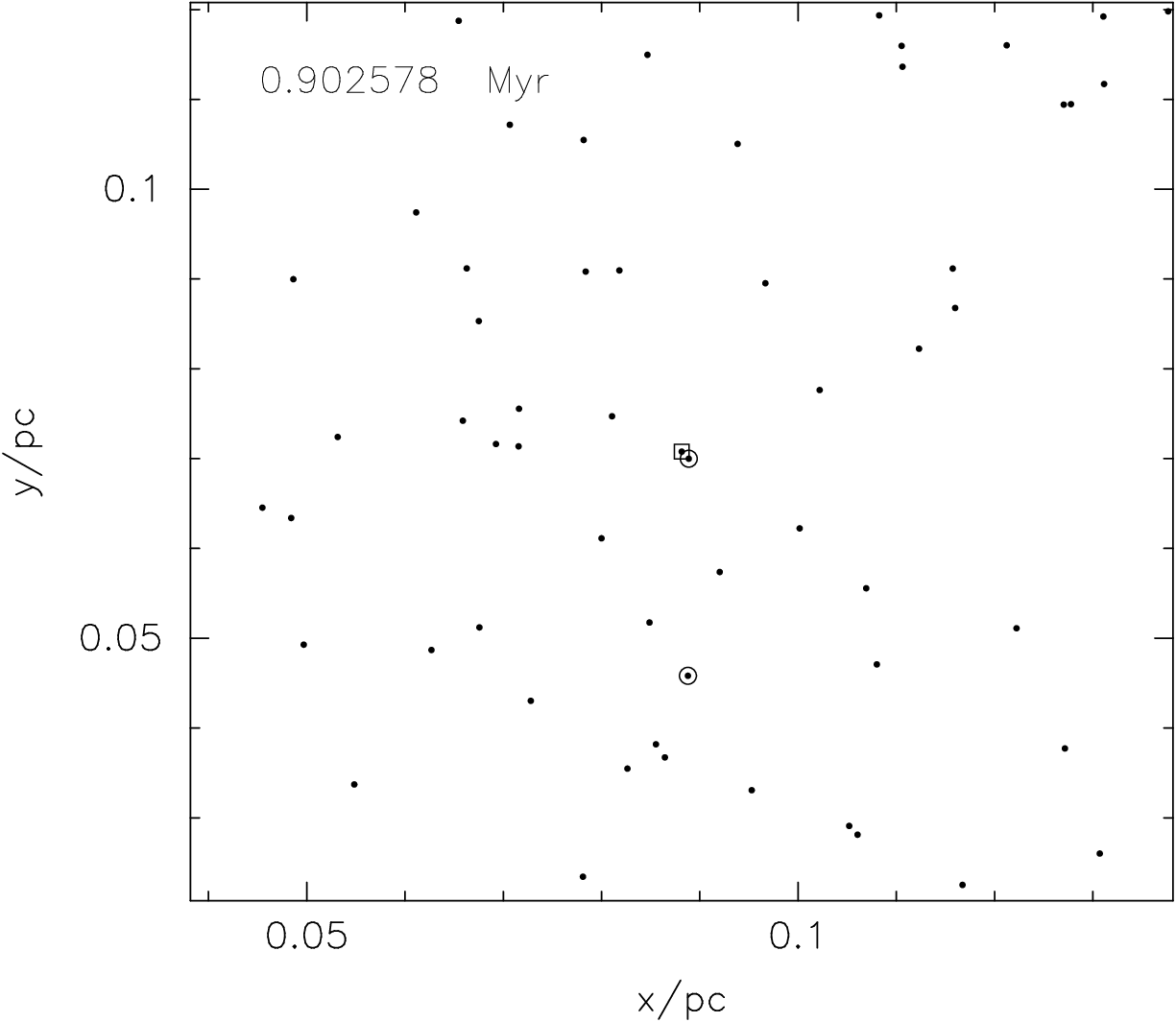}}
  \end{center}
  \caption[The trapezium systems detected in Run~a1.11]{Run a1.11:
    $D=1.6,Q=0.3$. {\it (a)} Members of trapezium systems for the run
    a1.11. The horizontal axis is measured in Myrs, and the vertical
    axis shows stellar mass. The points show the mass of the member
    stars in the detected trapezium systems, and different colours
    show different trapezium systems (Black: 41.3, 13.5, 11.9\Msun
    $\sim 0.7$ Myrs; Red: 41.3, 12.0, 9.9\Msun $\sim 1$ Myr; Green:
    41.3, 12.0, 8.7\Msun $\sim 1.8$ Myr; Blue: 41.3, 24.5, 12.0\Msun
    $\sim 1.8$ Myr).\\ {\it (b)} Projection of spatial positions for
    the first trapezium system (black) identified in {\it (a)}, at the
    time it was detected. {\it (c)} The members of the `black'
    trapezium shown at a time $\sim 0.1$~Myrs after its last
    detection, when it is not detected by our algorithm. It is
    clear that the original system has now evolved to a point where it
    no longer represents an `over-dense' group of neighbour
    stars. Trapezium system members are marked with outlines. The
    square outline identifies the most massive member.
  \label{fig:faa11a011_trap}}
\end{figure*}

We present here the details of the analysis of the trapezium systems
formed in a fairly typical cool, clumpy cluster in run a1.11 (with
$D=1.6$ and $Q=0.3$).  Figure~\ref{sfig:faa11a011.dat_trap} shows how
the existence and members of the trapezium systems in this cluster
evolve. Groups of points that occur at the same
time on the horizontal axis indicate that a trapezium system has been
identified at that particular time, and the points themselves show the
mass of the member stars. The different colours indicate the detection
of different trapezium systems, defined by a change in the three
highest mass stars in the system.

For example, our algorithm first finds a trapezium at $t\approx 0.7$
Myrs, which contains four stars of masses of around 41, 13, 12 and
2\Msun (black points). There are two important things to note about
this system.  Firstly, this trapezium system only lasts a very short
time, around 0.1~Myr, before decaying.  Secondly, the low-mass members
of the multiple system change even in this short time (e.g. a $\sim
5$~M$_\odot$ star is very briefly a member just before the system
decays). Figure~\ref{sfig:faa11a011.dat_trap.071} shows the spatial
distribution of this trapezium at the time of its first detection, the
trapezium members are identified by an outlined symbol. It is clear
from the figure that the quadruple system is at a higher spatial
density than the local stars and therefore should be detected as a
high-order multiple system, and because this system contains at least
three high-mass stars it is further defined as a trapezium
system. Figure~\ref{sfig:faa11a011.dat_trap.090} shows the system
$\sim 0.2$~Myrs later, when it is not detected by our algorithm. The
system has now dissolved, and is spread over a much larger area, thus
no longer satisfying the local density criteria.

A `new' trapezium (red points) then appears just after 1~Myr but lasts
for only one snapshot and contains two of the members of the first
trapezium.

Then just before 2~Myr a long-lived trapezium forms which survives for
the rest of the simulation (blue points).  This trapezium contains the
{\em same} $\sim 41$ and $\sim 12$~M$_\odot$ stars that were present in the first
and second systems, and also a new $\sim 24$~M$_\odot$ member.
Similarly to the first system to form this system contains several
lower-mass stars which come-and-go (for example a $\sim 10$~M$_\odot$
star is present for much of the first Myr of this system but is then
ejected). While the system is not detected at all times it has an almost
constant presence until the termination of the simulation, lasting at
least 2.2 Myrs.

\begin{figure*}
  \begin{center}
    \setlength{\subfigcapskip}{10pt}
\subfigure[]{\label{sfig:faa11a011.dat_trap.184b}
  \includegraphics[scale=0.35,angle=270]{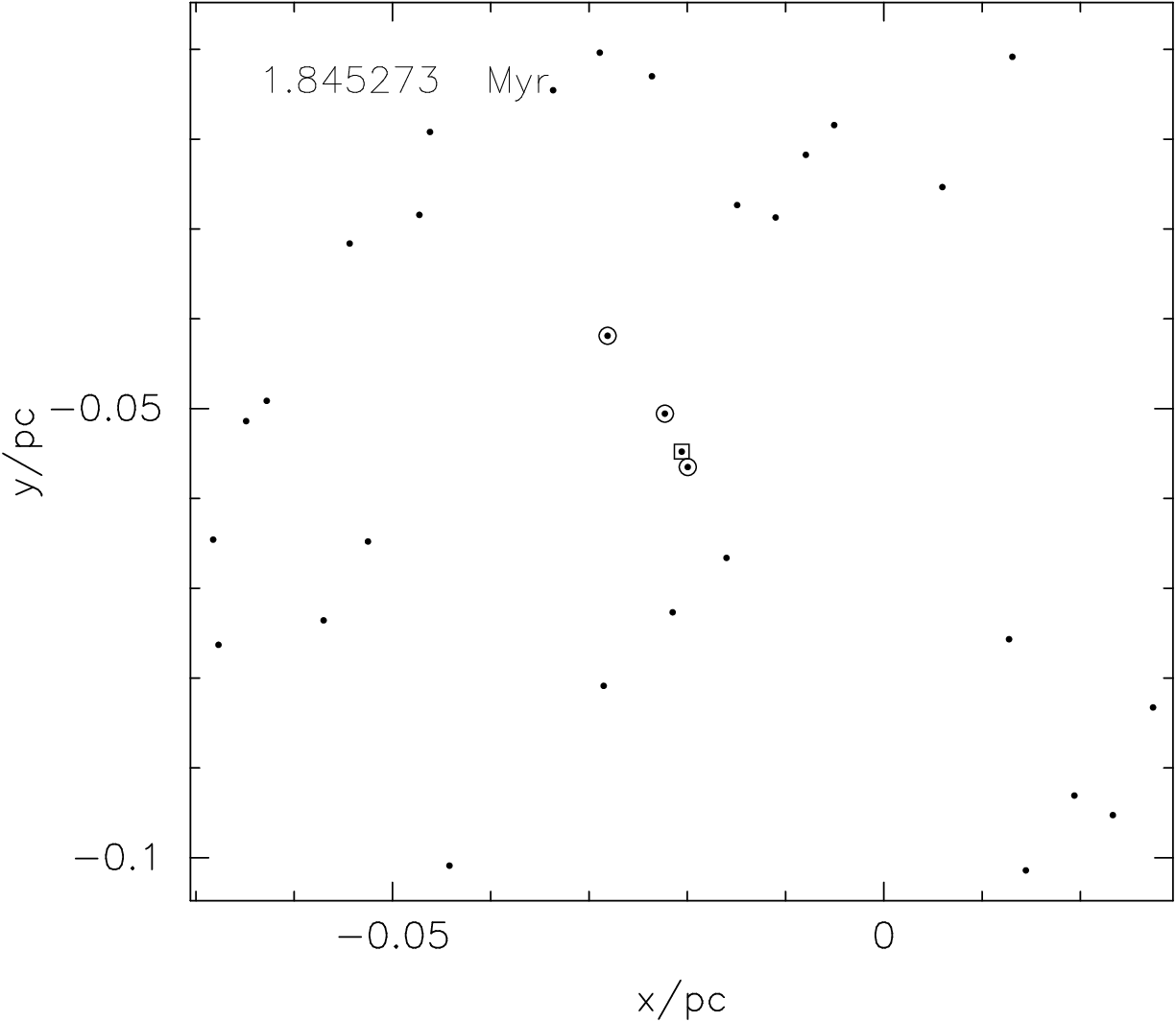}}
\subfigure[]{\label{sfig:faa11a011.dat_trap.193}
  \includegraphics[scale=0.35,angle=270]{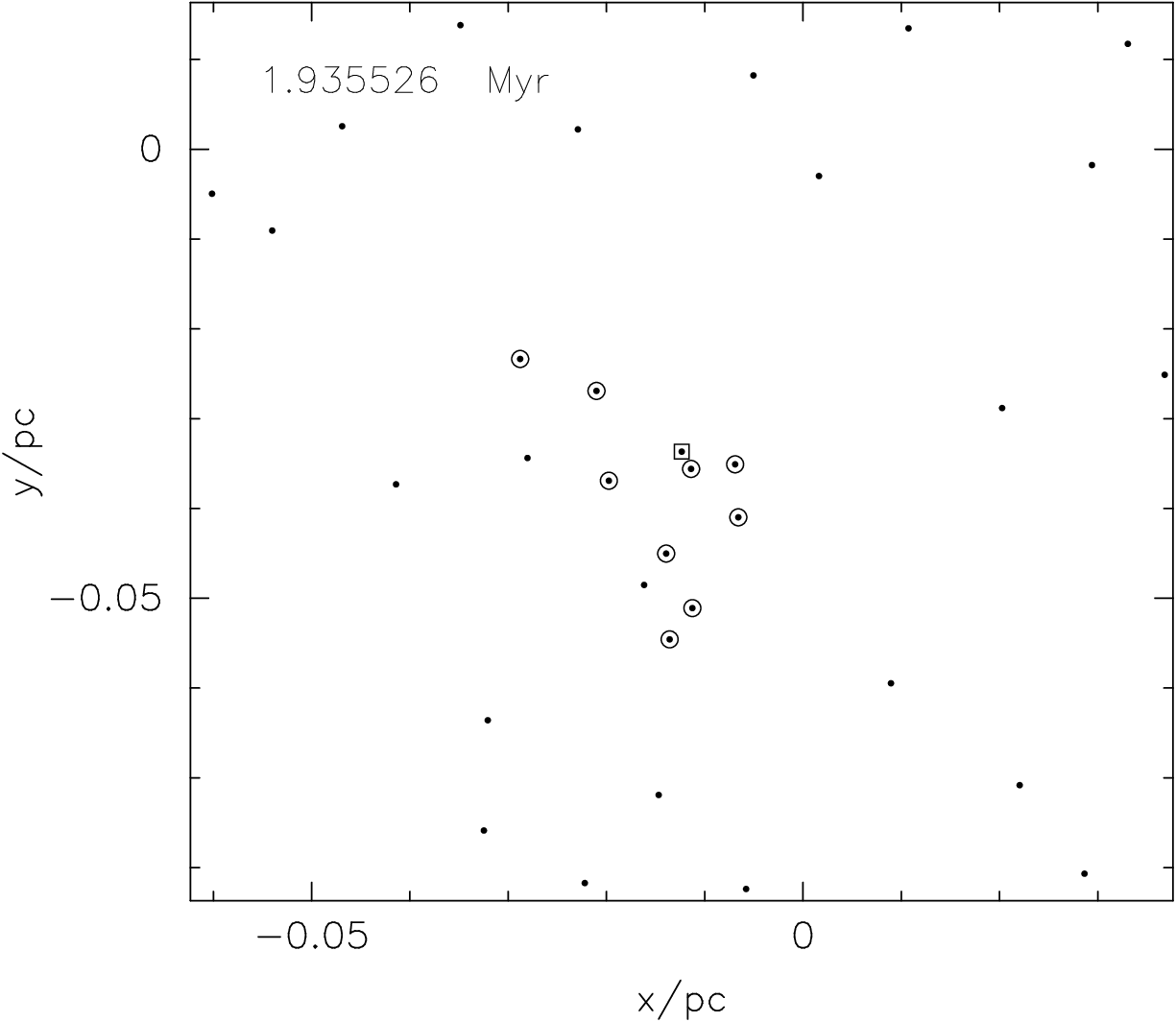}}
\subfigure[]{\label{sfig:faa11a011.dat_trap.232}
  \includegraphics[scale=0.35,angle=270]{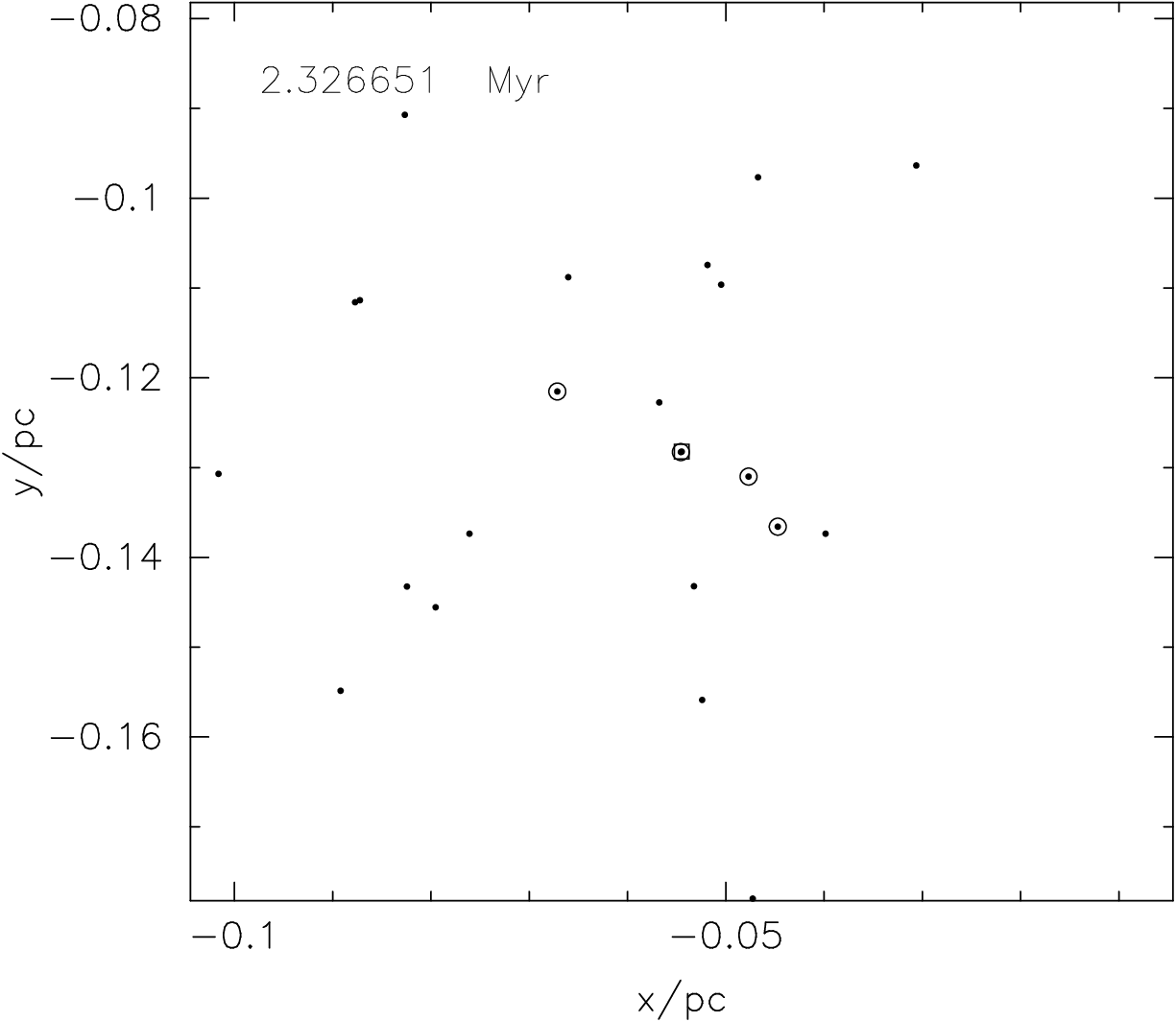}}
\subfigure[]{\label{sfig:faa11a011.dat_trap.264}
  \includegraphics[scale=0.35,angle=270]{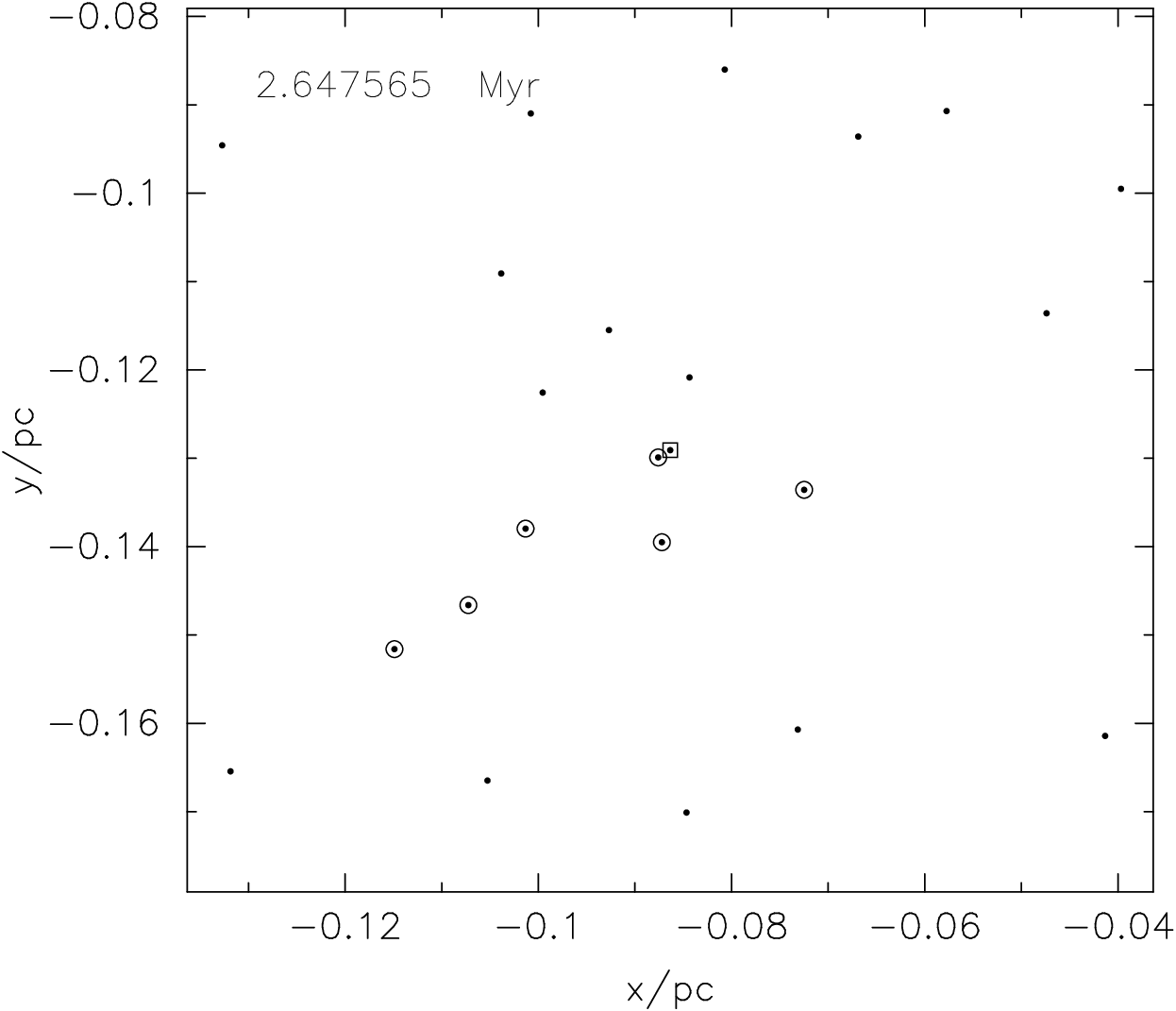}}
\subfigure[]{\label{sfig:faa11a011.dat_trap.300}
  \includegraphics[scale=0.35,angle=270]{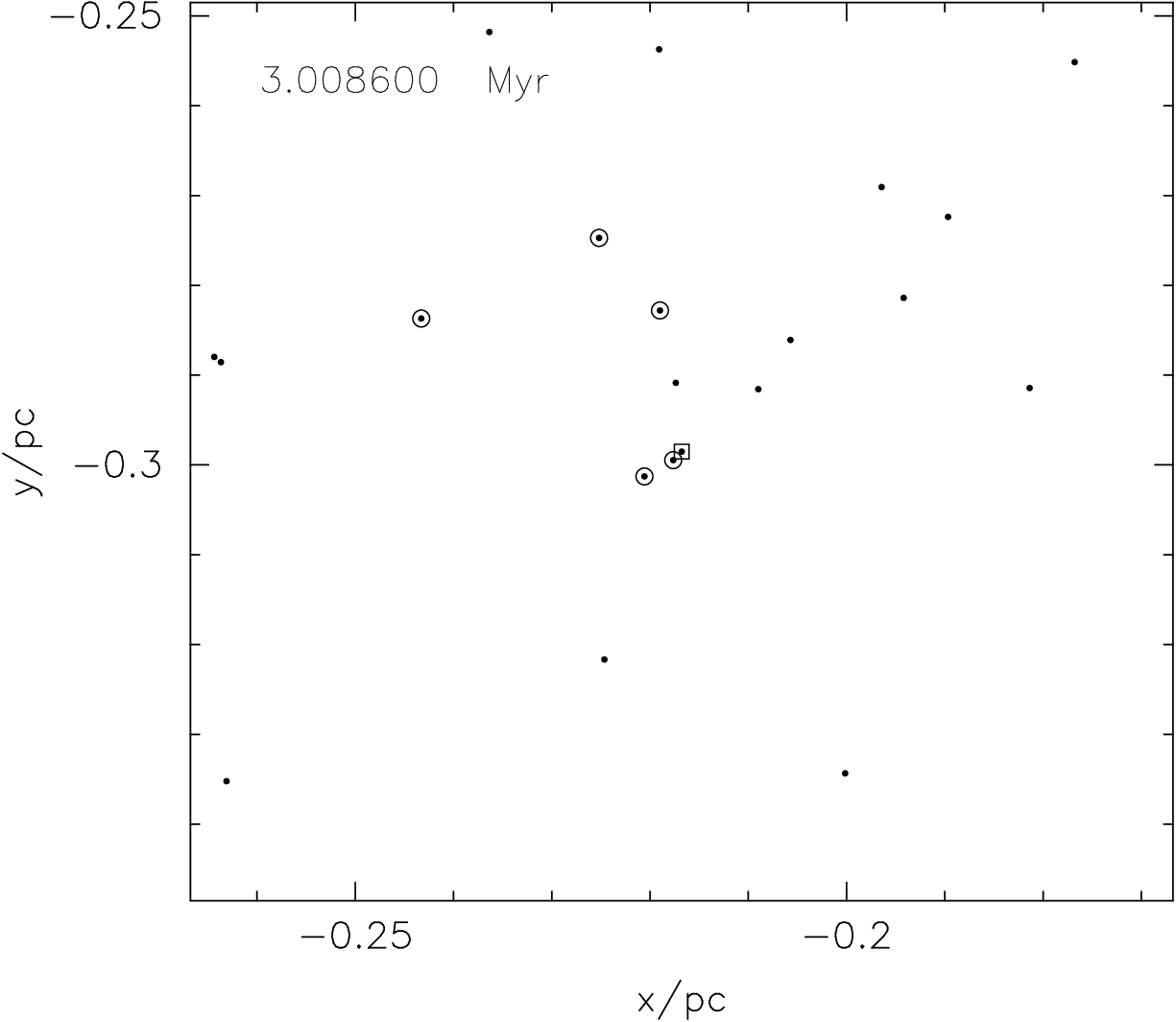}}
\subfigure[]{\label{sfig:faa11a011.dat_trap.399}
  \includegraphics[scale=0.35,angle=270]{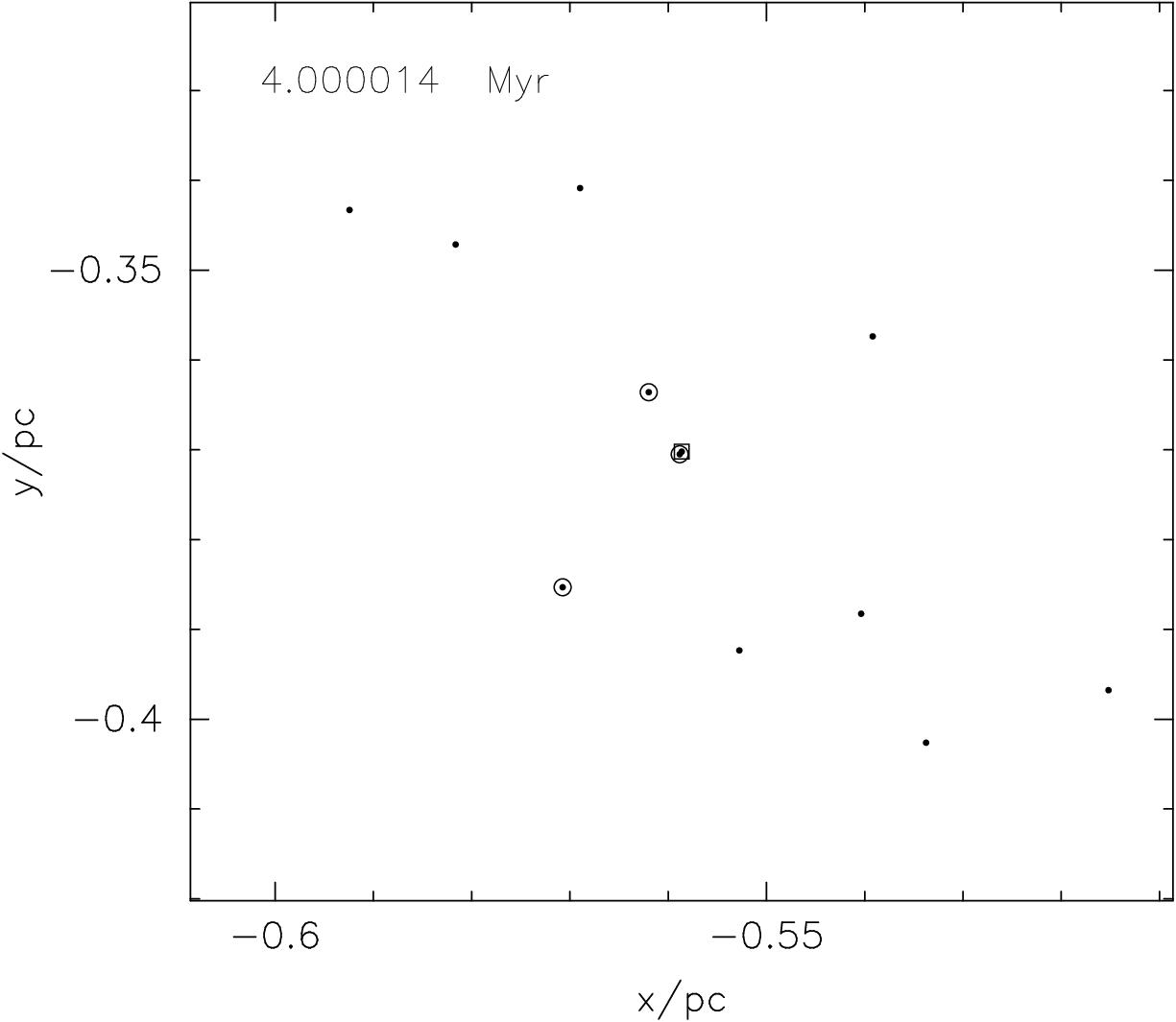}}
  \end{center}
  \caption[An in depth study of a trapezium system in Run~a1.11]{Run
    a1.11: $D=1.6,Q=0.3$.  Projection of spatial positions for the
    `blue' ($\sim$1.8 Myr) trapezium system identified in
    Figure~\ref{fig:faa11a011_trap}. Trapezium system members are
    identified marked with outlines. The square outline identifies the
    most massive member. The plots {\it (a)-(f)} show how the
    trapezium changes over the course of its life.
  \label{fig:faa11a011_bluetrap}}
\end{figure*}

Figure~\ref{fig:faa11a011_bluetrap} shows how the `blue' trapezium
identified in Figure~\ref{fig:faa11a011_trap} evolves spatially.  The
progression from Figure~\ref{sfig:faa11a011.dat_trap.184b} (1.8 Myrs)
to Figure~\ref{sfig:faa11a011.dat_trap.399} (4 Myrs), shows that the
trapezium system frequently adds and removes members, and that the
spatial distribution of the member stars changes often. For example in
Figure~\ref{sfig:faa11a011.dat_trap.193} the member stars are all
fairly evenly separated, with no obvious hierarchy; this is in
comparison to Figure~\ref{sfig:faa11a011.dat_trap.399}, in which a
hard binary (with the $\sim 42$\Msun and $\sim 12$\Msun stars at a
separation of  75~AU\footnote{For stars of this mass in a core with a
  velocity dispersion of $\sim 2$~km~s$^{-1}$ the hard-soft boundary
  is at around 1000~AU.}) is in a system with two other
stars. Trapezium systems have often been defined as `Hierarchical' or
`Trapezium' systems based on the separations of the member stars
\citep[e.g.,][]{allen74}. However, this maybe potentially misleading
as these systems can evolve between the states many times in their
lifetime.

The features shown in the simulation presented above are generic to
the simulations that produce trapezium systems, the supplementary
material contains figures similar to Figure~\ref{fig:faa11a011_trap}
for all of the simulations.  Examination of all of the simulations
shows that the initial conditions of the simulations do play a
significant role in the formation and evolution of trapezium systems.

\subsubsection{Trapezium frequency, lifetimes and sizes}
\label{ssec:trap_prop}


\begin{table}
  \centering
  \begin{tabular}{|c|c|c|c|c|}
\hline
        &  \multicolumn{4}{|c|}{$D$} \\
$Q$     &  1.6    &  2.0   &  2.6   &  3.0   \\
\hline
0.3     &  45/50  &  4/10  &  2/10  &  4/10  \\
0.4     &  32/50   &  8/10  &  4/10  &  1/10  \\
0.5     &  5/10   &  5/10  &  0/10  &  1/10  \\
\hline
  \end{tabular}
\caption[The number of clusters that contain trapezium systems]{The
  number of simulations which produce a trapezium system and its
  dependence on the virial ratio $Q$ and fractal dimension $D$ of the
  initial cluster.
\label{tab:contain_trap}}
\end{table}


Table~\ref{tab:contain_trap} shows the fraction of simulations that
form a trapezium system (at least for one snapshot) and its dependence
on the the virial ratio, $Q$, and fractal dimension, $D$, of the
initial cluster.  An analysis of Table~\ref{tab:contain_trap} shows
that trapezium systems are much more likely to be found in clusters
that initially have more substructure and are cooler. As we look
across the table, from more ($D=1.6$) to less ($D=3.0$) substructured,
we can see that there is a drop in the number of systems that show a
trapezium system beyond $D=2.0$. Two-thirds of the clusters with
$D\leq 2.0$ form a trapezium system, compared to one-fifth with
$D>2.0$. The same trend is seen as we look down the table from cooler
($Q=0.3$) to virialised clusters, although this effect is much less
pronounced. Clusters that are initially cool and substructured
collapse to a much denser state, allowing the massive stars to mass
segregate and form trapezium systems \citep{allison09b,allison10}.


\begin{table}
  \centering
  \begin{tabular}{|c|c|c|c|c|}
\hline
        &  \multicolumn{4}{|c|}{$D$} \\
$Q$     & 1.6  & 2.0  & 2.6  & 3.0   \\
\hline
0.3     & 1.52 & 2.13 & 2.50 & 3.13  \\
0.4     & 2.00 & 2.13 & 3.63 & 3.25$^\star$  \\
0.5     & 2.05 & 3.45 & -    & 3.25$^\star$  \\
\hline
  \end{tabular}
\caption[The average formation time of trapezium systems]{The average
  formation time for trapezium systems in Myr in simulations with
  different fractal dimensions ($D$) and virial ratios ($Q$).  Note
  that these averages are over the simulations which form trapezium
  systems, and in the $D=3.0$, $Q=0.4$ and $0.5$ ensembles marked by a
  $^\star$ only one system forms and so this is the formation time of
  those single systems rather than an average.  
\label{tab:form_trap}}
\end{table}


At what age trapezium systems form is also dependent on the initial
conditions of the simulation. If we define the first detection of a
trapezium system in a particular run as the `formation time', we find
that in clusters that are initially cool and more substructured
trapezium systems tend to form at earlier times. In the lowest-$Q$ and
$-D$ clusters trapeziums tend to form early ($< 2$~Myr), whilst in
high-$Q$ and high-$D$ clusters trapeziums form close to the end of the
simulations at around $3$ -- $4$~Myr (the simulations end at
$4$~Myr). This can be explained by the initial collapse of the cluster
-- clusters that are initially further from an equilibrium state go
through a more dramatic violent relaxation event. The depth of the
collapse in these clusters allows the massive stars to encounter each
other much sooner in the evolution of the cluster, compared to
clusters that are initially close to equilibrium.

\begin{figure}
  \begin{center}
    \setlength{\subfigcapskip}{10pt}
\subfigure[a1]{\label{sfig:trap_per_time_a1}
  \includegraphics[scale=0.35,angle=270]{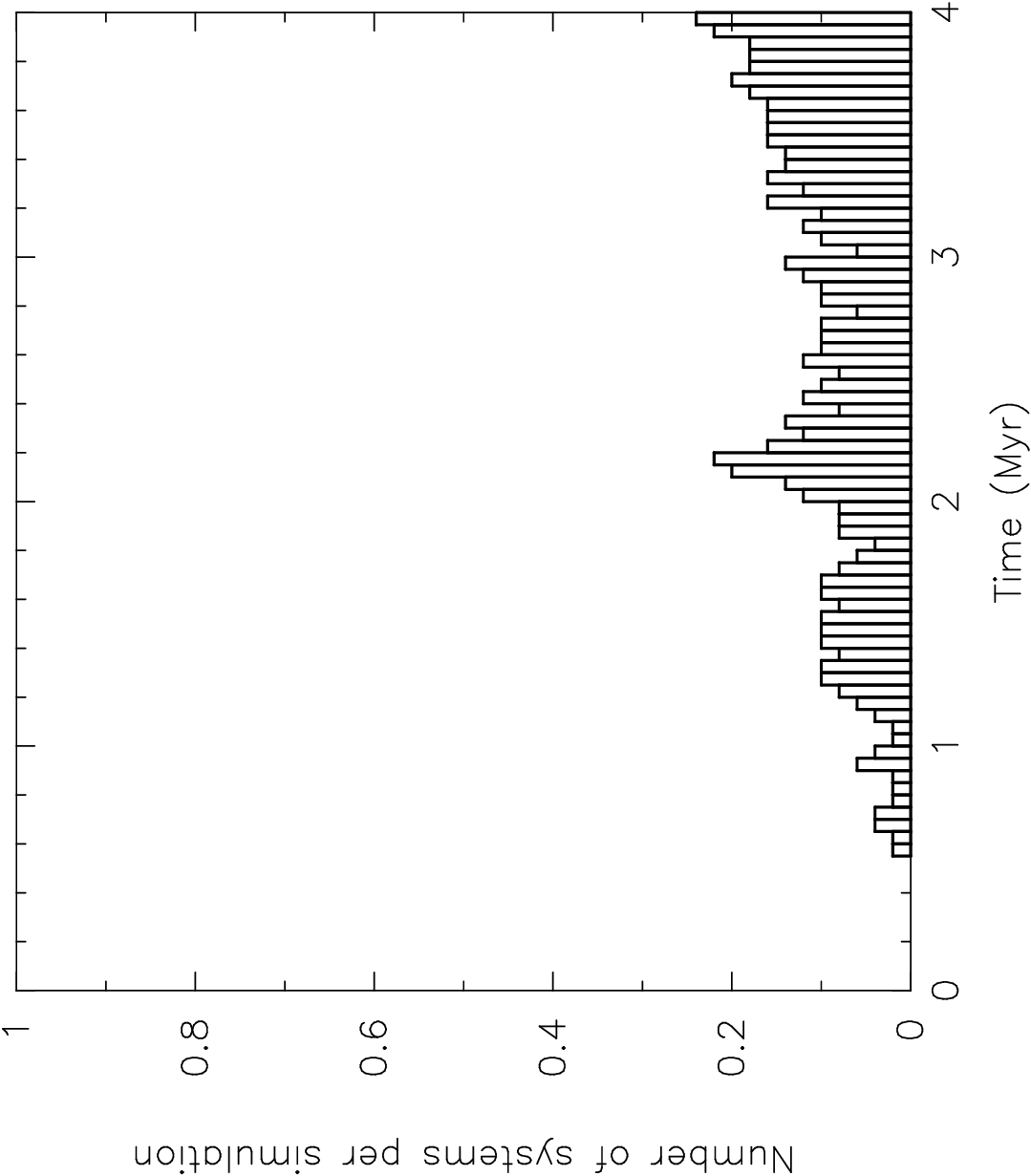}}
\subfigure[b1]{\label{sfig:trap_per_time_b1}
  \includegraphics[scale=0.35,angle=270]{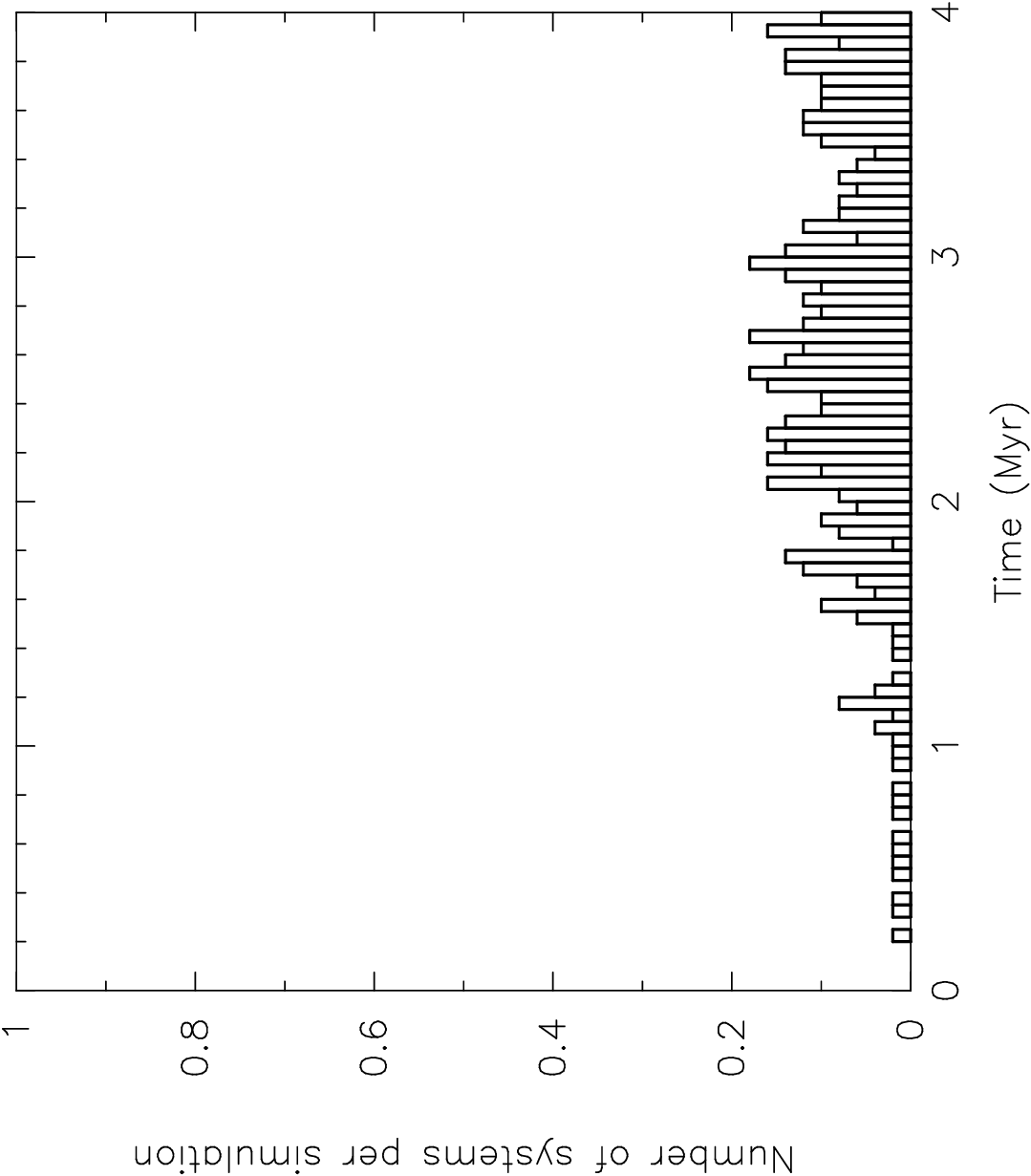}}
  \end{center}
  \caption[]{Plots show the fraction of clusters with trapezium-like
    systems changes as a function of time in the
    `a1-' ({\it top}) and `b1-type' ({\it bottom}) simulations. The
    plots for all simulations types can be found in the Supplementary
    Material.
  \label{fig:trap_per_time}}
\end{figure}

The length of time a trapezium can be `observed' in any cluster also
depends on the initial conditions.  Note that as seen above each
cluster may have several different trapeziums with different members,
here we discuss the time over which {\em any} trapezium is in
existence. Figure~\ref{fig:trap_per_time} shows how likely it is that
a trapezium-like system will be observed (i.e. is detected by our
algorithm) in a particular cluster-type at any time during the life of
those clusters.  
This figure shows the plots for the `a1-' ($Q=0.3$, $D=1.6$, 
{\it top}) and `b1-type'
($Q=0.4$, $D=1.6$, {\it bottom}) clusters, the plots for all the 
simulations can be found in the Supplementary Material. 

The low numbers of high-$D$ and high-$Q$ clusters found to have
trapeziums in Table~\ref{tab:contain_trap} (typically 1 or 0) is
partly due to the longevity of their trapeziums.  Therefore the
fractions found in Table~\ref{tab:contain_trap} may be unreliable as
there may be very short-lived trapeziums that exist between snapshots
that we miss.  However, if this were to occur it would still be
extremely unlikely to observe a trapezium in such clusters as their
lifetimes even if they do form are extremely short ($< 0.1$~Myrs).


\begin{table}
  \centering
  \begin{tabular}{|c|c|c|c|c|}
\hline
        &  \multicolumn{4}{|c|}{$D$} \\
$Q$     & 1.6  & 2.0  & 2.6  & 3.0   \\
\hline
0.3     & 3.7  & 3.8  & 2.5  & 2.0  \\
0.4     & 2.0  & 2.6  & 1.8  & 1.0$^\star$ \\
0.5     & 2.4  & 1.4  & -    & 1.0$^\star$  \\
\hline
  \end{tabular}
\caption[The average number of trapezium systems]{Average number of
  different trapezium systems present in simulations with different
  fractal dimensions ($D$) and virial ratios ($Q$).  Again note
  that these averages are over the simulations which form trapezium
  systems, and the ensembles
  marked with a $^\star$ have only one system that briefly appears and
  so do not represent an average.
\label{tab:sysnum_trap}}
\end{table}


This point is emphasised in Table~\ref{tab:sysnum_trap} which shows
the different numbers of trapezium systems typically found in different
simulations.  We identify different trapezium systems by a change in
the three most massive members.  The dynamical nature of many
trapezium systems is shown here as the decay, re-formation and
swapping of members can frequently significantly change the trapezium
system observed in any one cluster. We should note that the longevity
of the trapeziums in our simulations is limited by the 4~Myr duration
of the simulations.


  \begin{table}
    \centering
    \begin{tabular}{|c|r|c|l|c|}
      \hline
      $N$  & \multicolumn{3}{c}{Length (AU)}  & \# systems   \\
      \hline
      3    & 5143  &  $\pm$ & 469   & 776         \\
      4    & 9575  &  $\pm$ & 1085  & 739         \\
      5    & 11279 &  $\pm$ & 1157  & 418         \\
      6    & 16169 &  $\pm$ & 2346  & 196         \\
      7    & 16890 &  $\pm$ & 2290  & 70          \\
      8    & 18864 &  $\pm$ & 2253  & 23          \\
      9    & 18613 &  $\pm$ & 3085  & 4           \\
      \hline
    \end{tabular}
    \caption[The average size of trapezium systems]{The average MST
      size of all trapezium systems with $N$ members in all
      simulations with 1$\sigma$ errors, as well as the total number 
      of snapshots in which
      systems of each $N$ are found.
  \label{tab:len_trap}}
  \end{table}


The size of a trapezium is a rather difficult property to quantify.
We define the size of a trapezium to be the length of the minimum
spanning tree (MST)\footnote{The MST is the shortest path which
  connects all of the vertices in a sample with simple edges, and no
  closed loops \citep{prim57}} connecting all of the members.  This
provides a unique length with which the size of the trapeziums can be
described. For reference, the Trapezium system in the ONC is an
$N\approx 8$ system, and has an MST length $\approx 13600$ AU
\citep[assuming a distance of 440~pc;][]{jeffries07}.  In
Table~\ref{tab:len_trap} we show the average MST lengths for all
$N=3-9$ systems.  This data was taken for every snapshot of every
simulation as the size and membership of trapeziums changes even
within a single simulation.  As can be seen the ONC Trapezium is
smaller than an average $N\approx 8$ system, although this is likely
due to the fact that the Trapezium system in the ONC contains a
`mini-trapezium', which is included in this MST analysis. The
hierarchical nature of the Trapezium probably occurs due to the
initial primordial binary population of the ONC. Primordial binaries
are not included in these simulations, which probably explains why we
do not see the levels of hierarchy seen the the ONC Trapezium. 
  The ONC Trapezium could also be smaller than the average lengths
  presented here because it is currently in a dense phase of dynamical
  evolution. One of the main conclusions of this work is that
  trapezium-like systems are extremely dynamical entities.

\begin{figure}
  \begin{center}
    \setlength{\subfigcapskip}{10pt}
\includegraphics[scale=0.3,angle=270]{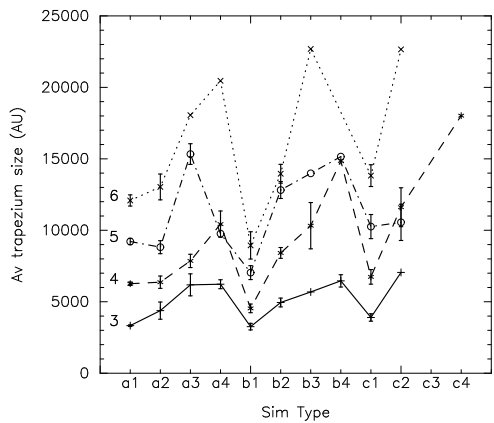}
  \end{center}
  \caption[The average size of trapezium systems]{The average size of
    trapeziums of $N=3,4,5$ and $6$ (bottom to top, indicated by
    numbers on left of plot), for the different initial conditions
    letters correspond to initial virial ratio a: $Q=0.3$, b: $Q=0.4$,
    and c: $Q=0.5$, and numbers to the initial fractal dimension 1:
    $D=1.6$, 2: $D=2.0$, 3: $D=2.6$, and 4: $D=3.0$. Error bars show
    standard error.
  \label{fig:trap_size}}
\end{figure}

Figure~\ref{fig:trap_size} shows how the average size of a trapezium
system changes with the initial conditions of the simulation.  To
recap Table.~\ref{tab:runs}, simulation letters correspond to initial
virial ratio a: $Q=0.3$, b: $Q=0.4$, and c: $Q=0.5$, and numbers to
the initial fractal dimension 1: $D=1.6$, 2: $D=2.0$, 3: $D=2.6$, and
4: $D=3.0$.  Figure~\ref{fig:trap_size} shows that trapeziums tend to
be smaller when clusters are initially low-$D$ and low-$Q$ (e.g. a1
and b1), and larger when initially high-$D$ and high-$Q$ (e.g. c3, b4
and c4). However, it is interesting to note that clusters that
  begin as `b1-type' ($D=2.0,Q=0.4$) seem to have the smallest
  trapezium systems, which may indicate that these initial parameters
  provide a good environment to form tightly bound high-order multiple
  systems. As is shown in
Tables~\ref{tab:contain_trap}~and~\ref{tab:sysnum_trap} clusters that
begin with warmer and smoother initial distributions form only a few
trapezium systems. Therefore, the average size of trapeziums in these
clusters is statistically less reliable than in the cooler, clumpier
clusters.

\subsection{The formation of trapeziums}
\label{ssec:formation}

We have seen that in initially low-$D$ and low-$Q$ clusters many small
trapeziums are formed, they tend to form early ($<2$~Myrs), and can
survive for a significant fraction of the cluster's early life ($\sim
2$~Myrs).  However, in initially high-$D$ and high-$Q$ clusters far
fewer and larger trapeziums form, they form later, and they have
shorter lifetimes.  We have also seen that trapeziums are highly
dynamic entities that constantly change their low-mass members and
also decay, re-form, and swap higher-mass members.

These properties come about because in our simulations
trapeziums are most often formed during the dense phase of the
collapse of a young star cluster.  As described in detail in
\citet{allison09b,allison10} clusters which are out-of-equilibrium
undergo a rapid ($<1$~Myr) collapse and violent relaxation phase as
they attempt to virialise.  In clusters with a low-$Q$ and low-$D$
their radii may shrink initially by a factor of several leading to a
short-lived, but extremely dense phase.  In this phase the two-body
relaxation time can be extremely short (i.e. $<1$~Myr) and clusters
can dynamically mass segregate their most massive stars (in the case
of Orion-like clusters this is stars more massive than a few Solar
masses).

This leads to a situation where the most massive stars are centrally
concentrated and in which the formation of trapezium systems is
relatively easy.  The densest phase is usually reached at around 1~Myr
(about an initial free-fall time) explaining why trapeziums in low-$Q$ and
low-$D$ clusters tend to first form on such a timescale.

The higher both $D$ and $Q$ become the deeper the depth of the collapse
and so the lower the degree of mass segregation and trapezium
formation.  Low-$D$ is the dominant factor as the degree of collapse
is more sensitive to this parameter than the virial ratio $Q$.  This
is because initially clumpy distributions have a larger initial net
potential energy ($|\Omega|$), which increases during the collapse of
the cluster, therefore allowing the collapse to form a smaller final
cluster compared to less clumpy initial conditions. In initially
virialised and smooth clusters there may be some collapse and
relaxation (especially as the cluster moves from initially uniform
density to a more Plummer-like density distribution), but it is far
less extreme and violent.  Trapeziums that form in such clusters are
freak events and tend to decay on timescales of a few of their own
crossing times.

Formation during collapse also explains the difference in the size of
trapezium systems.  In low-$Q$ and low-$D$ clusters, the collapse is
deeper producing smaller trapeziums, whilst in high-$Q$ and
high-$D$ clusters if a trapezium does form it is more by chance and
will tend to be much larger as the chance of a very close encounter
between several of the most massive stars is fairly low.

\subsubsection{The dissolution of trapezium systems and their host clusters}
\label{ssec:diss_trap}

We have shown that trapezium systems are frequently formed through
dynamical interactions in cool and clumpy clusters, but how and at
what ages would we expect these systems to decay? There are three main
decay paths for the massive star multiples that are formed in our
simulations: the supernova of the most massive member; the dynamical
decay of the trapezium system itself possibly, destroying the host
cluster in the process; and the dissolution of the cluster around the
system (most likely caused by gas expulsion) leaving an unstable
few-body system that will dynamically decay in a few crossing times.

The dynamical decay of trapezium systems occurs due to the inherent
instability of the few-body multiple systems. The decay occurs when at
least three of the massive stars in the system approach closely to
each other. The encounter leads to the ejection of massive stars from
the system, leaving a hard massive-star binary. The binary can
increase its binding energy by as much as the entire potential energy
of the cluster. The decay of multiple systems does not appear to occur with
any regularity or obvious trigger, indicating that this decay
mechanism is chaotic. It is dependant only on the chance encounter
between massive stars.

Dynamically decaying trapezium systems can lead to the dissolution of
the host clusters they reside in; but the converse is also true -- the
dissolution of a cluster around a trapezium can lead to the decay of
the massive multiple. The cluster can dissolve around the
multiple system because of a rapid change in the cluster's potential
due to the expulsion of gas -- the group of luminous massive stars in
the core will most likely drive out the natal gas from the
cluster. When this occurs the rapid change in potential causes the
cluster to attempt to re-virialise itself, leading the cluster to
 rapidly expand on a timescale of $\sim 2-5$~Myr
  \citep{goodwin06}.  This will leave the trapezium as a lone, and
  highly unstable, high-order multiple with no host cluster to replace
  ejected stars.

If these other mechanisms have not destroyed the trapezium system by
the time the most massive member becomes a supernova, the trapezium
system could be destroyed by the supernova of the most massive
member. The supernova will cause a huge loss of mass from the system
(and quite possibly eject the remnant due to a large kick velocity).
Obviously, the timescale that this decay path would occur on is
dependant on the mass of the most massive member.  Even if the
supernova of the most massive member is not catastrophically
destructive, all of the massive stars in a trapezium will go
supernovae within a few tens of Myr at most meaning that the system is
no longer a trapezium (as it has no massive stars).

\section{Discussion}
\label{sec:discussion}

We have analysed $N$-body simulations of initially cool and clumpy
star clusters, and of the formation and evolution of
trapezium-like systems. We find that trapezium system formation is
dependant on the initial conditions of the host cluster -- with
clumpier and cooler clusters more likely to form trapezium systems. 
We also find that the physical properties of trapezium systems,
such as size, are also dependant on the initial state of the cluster. Importantly we find that 
that trapezium systems are highly dynamical, and often transient, 
objects that add, remove and swap members often.

The formation of trapezium systems is an almost ubiquitous process in
the collapse of cool and clumpy clusters, with around 75 per cent of clusters
with $Q=0.3-0.4$ and $D=1.6-2.0$ forming a trapezium system during the
first 4 Myrs of their dynamical evolution. In contrast, $<10$ per cent of
clusters with relatively little substructure
($D=2.6-3.0$) and close to virial equilibrium form trapezium
systems. From this analysis, if the formation of stars is a cool and
clumpy process we should expect to see trapezium systems in many young
clusters ($\sim$2 -- 4~Myr), that have recently aggregated into a
newly formed `cluster' \citep{zinnecker07,zinnecker08}.  The
  trapezium formation process described in this work and by Zinnecker
  are qualitatively similar to each other in that both processes
  involve the merger of a clumpy stellar distribution, which leads to
  the formation of a trapezium system. The main difference between the models
  is that Zinnecker proposes that the clumpy distribution of stars is
  one in which there are distinct subclusters, each containing
  a massive star. The merger of these subclusters then brings the
  massive stars together and a trapezium system is formed. The process
  described in this work instead proposes a hierarchical distribution
  of stars, and a random distribution of massive stars, and it is
  this global collapse and subsequent dynamical mass segregation 
which brings the massive stars together. Both
  models propose that trapezium systems would be formed by the
  dynamical evolution of the merging cluster stars.

The constant changing of trapezium systems (e.g. 
Figure~\ref{fig:faa11a011_bluetrap}) suggest that
trapezium systems are highly dynamical objects. From this analysis, it
appears very likely that the trapezium system in the ONC that we
observe today is only a snapshot in its lifetime; {\it the system very
  probably appeared different in the past, and will likely change in
  the future} by incorporating new members into the system, discarding
current members, changing its spatial distribution and eventually
dissolving -- possibly destroying the ONC at the same time. Recent
investigations have in fact already indicated that the ONC has ejected
some member stars \citep{tan04,poveda05,pflamm-altenburg06}.

Observations of the `canonical' trapezium system located in the ONC
find that it is hierarchical; the member stars in the Trapezium system
are all multiple, found to be in either binary systems or in
`mini-trapeziums' of their own
\citep{preibisch99,kraus07,moeckel09b}. It should be mentioned here
that we do not form such hierarchical systems, and usually find that
only the most massive stars in the trapezium systems that are formed
appear to be in lasting binary systems. As we begin our simulations
with only single stars this may not be surprising, but the result does
show that to produce trapezium systems with the multiplicity seen in
the ONC primordial multiplicity is important.

Observations of two high-mass star forming regions have found possible
examples of trapezium systems in embedded clusters. \citet{megeath05}
find a deeply embedded $N=5$ system in W3. This system has a maximum
projected separation between the five sources of $\sim 5600$~AU, which
would likely indicate that this system would be detected as a
trapezium system by our algorithm. Observations of the star forming
region NGC~7538 find a possible system consisting of nine
submillimeter cores, with masses between $20\pm 11 - 6\pm 3$\Msun
within a 0.3$\times$0.3~pc area \citep{qiu10}. This system has a
large MST length ($\sim 160,000$~AU), and so is substantially larger
than the $N=9$ systems we find in our simulations (which have a MST
length $\sim 18,000\pm 3,000$~AU) and is therefore unlikely to be
classified as a trapezium system though our definition. However,
because of the age of this cluster, the potential system could be in
the process of forming. Kinematical information would be of great use
for determining the future evolution of this system.

\section{Conclusions}
\label{sec:tconc}

We have simulated the early $N$-body dynamical evolution of Orion-like
($N=1000$) star clusters with a range of initial conditions from cool
and clumpy to smooth and relatively relaxed.  We find that the
formation of higher-order massive star multiples similar to the
Trapezium in the ONC to be fairly common in cool and clumpy star
clusters, but rare in those that start smooth and relatively relaxed.

Our main conclusions can be summarised as:

\noindent $\bullet$ High-order massive multiple (trapezium) systems 
can form dynamically during the dense collapse of cool, clumpy
clusters.  {\em Therefore trapezium systems have no need to be primordial.}

\noindent $\bullet$ Trapezium systems change their size, structure,
and membership (especially low-mass) frequently with trapeziums rarely
looking the same over timescales of only a few $\times 10^5$~yrs.
{\em Therefore drawing conclusions from the current appearance of a
  trapezium system must be done with great care.}

\noindent $\bullet$ Trapeziums can be very long-lived (up to 2 --
3~Myr in our simulations that only last for 4~Myr) only in the sense that a
trapezium system is observable, however the dynamical nature of the
trapeziums means it is difficult (to say the least) to say that the
`same' trapezium is always present.

Like the Trapezium in the ONC, the trapezium systems that we find
have, by definition, at least three massive ($> 8$~M$_\odot$) stars,
but also generally have several low-mass members as well. However,
unlike the Trapezium system, our trapeziums do not display the close
binaries and `mini-trapeziums' that are observed. As the initial
conditions of the simulations do not include primordial binaries, this
is a likely reason why we do not see such a phenomenon.

\section{Acknowledgements}

We would like to thank David Hubber for many useful conversations, and
the anonymous referee for a thorough report and comments that have
improved the paper. RA acknowledges financial support from STFC and
the University of Sheffield. This work has made use of the {\sc
  Iceberg} computing facility, part of the White Rose Grid computing
facilities at the University of Sheffield.

\bibliography{2552bib}
\bibliographystyle{mn2e}

\end{document}